\documentclass[conference]{IEEEtran}
\IEEEoverridecommandlockouts
\usepackage{cite}
\usepackage{amsmath,amssymb,amsfonts}
\usepackage{algorithmic}
\usepackage{graphicx}
\usepackage{textcomp}
\usepackage{xcolor}
\usepackage{subfigure}
\usepackage{dblfloatfix}
\usepackage{lipsum}
\def\BibTeX{{\rm B\kern-.05em{\sc i\kern-.025em b}\kern-.08em
    T\kern-.1667em\lower.7ex\hbox{E}\kern-.125emX}}
\begin{document}

\title{Live Visualization of Dynamic Software Cities with Heat Map Overlays}

\author{\IEEEauthorblockN{Alexander Krause}
\IEEEauthorblockA{\textit{Department of Computer Science} \\
\textit{Kiel University}, Kiel, Germany \\
akr@informatik.uni-kiel.de}
\and
\IEEEauthorblockN{Malte Hansen}
\IEEEauthorblockA{\textit{Department of Computer Science} \\
\textit{Kiel University}, Kiel, Germany \\
malte.hansen@stu.uni-kiel.de}
\and
\IEEEauthorblockN{Wilhelm Hasselbring}
\IEEEauthorblockA{\textit{Department of Computer Science} \\
\textit{Kiel University}, Kiel, Germany \\
wha@informatik.uni-kiel.de}
}

\maketitle

\begin{abstract}
The 3D city metaphor in software visualization is a well-explored rendering method.
Numerous tools use their custom variation to visualize offline-analyzed data.
Heat map overlays are one of these variants.
They introduce a separate information layer in addition to the software city's own semantics. 
Results show that their usage facilitates program comprehension.

In this paper, we present our heat map approach for the city metaphor visualization based on live trace analysis.
In comparison to previous approaches, our implementation uses live dynamic analysis of a software system's runtime behavior.
At any time, users can toggle the heat map feature and choose which runtime-dependent metric the heat map should visualize.
Our approach continuously and automatically renders both software cities and heat maps.
It does not require a manual or semi-automatic generation of heat maps and seamlessly blends into the overall software visualization. 
We implemented this approach in our web-based tool ExplorViz, such that the heat map overlay is also available in our augmented reality environment.
ExplorViz is developed as open source software and is continuously published via Docker images.
A live demo of ExplorViz is publicly available.
\end{abstract}


\section{Introduction}\label{sec:introduction}

Introducing new technologies, code refactorings, or new developers into a software system's development project requires significant effort.
A reason is that program comprehension is a time consuming task in software development~\cite{Bennett2002}.
Professional developers use integrated development environments, web browsers, and document editors to facilitate common tasks in the context of program comprehension~\cite{xia2018}.
The software visualization research community introduces various methods and techniques to render the structure, evolution, or runtime behavior of software systems, therefore, aid developers to understand their systems.

The 3D city metaphor~\cite{wettel2007} is a popular visualization method to render software applications as cities.
In the context of object-oriented languages, these software cities depict classes as three-dimensional boxes that are grouped by surrounding tiles, i.e., source code packages.
While the city metaphor is mainly used in the context of static code analysis~\cite{jeffery2019,pfahler2020}, there are also approaches that focus on dynamic analysis and its resulting visualization to incorporate software dynamics~\cite{weninger2020,hasselbring2020}, such as a software system's runtime behavior.

Benomar et al.~\cite{benomar2013} present a variant of the city metaphor visualization that is extended by heat maps.
Their approach visualizes the contribution of developers to source code classes by overlaying software cities with heat maps.
In addition, they recorded and pre-processed execution traces of their sample application to visualize the level of activity for each class during execution.

Inspired by this work, we present an approach that continuously and automatically renders both software cities and heat maps, in near realtime within one tool~\cite{Reck2020}.
Our implementation uses live dynamic analysis of a software system's runtime behavior.
The heat map overlay visualizes live calculated scores of user-selected metrics.
It does not require a manual or semi-automatic generation of heat maps.
We implemented this approach in our web-based open source tool ExplorViz~\cite{fittkau2017,hasselbring2020}.
A live demo is online available.\footnote{https://www.explorviz.net}
The heat map feature can also be used in our augmented reality (AR) environment.
This extension will also be introduced in our virtual reality (VR) mode~\cite{Fittkau2015}.

\section{Related Work}\label{sec:related-work}
Benomar et al.~\cite{benomar2013} present a heat map approach for the visualization
of software evolution and execution traces.
Their approach is built on top of the framework VERSO~\cite{langelier2005}.
VERSO uses a treemap layout to display package structures as hierarchically organized regions and classes
as three-dimensional boxes within those regions. Statically calculated metric scores, for instance concerning complexity,
can be mapped to different visual properties of the class representations, e.g. their color or rotation.

\begin{figure*}[!b]
	\centering
	\includegraphics[width=\textwidth]{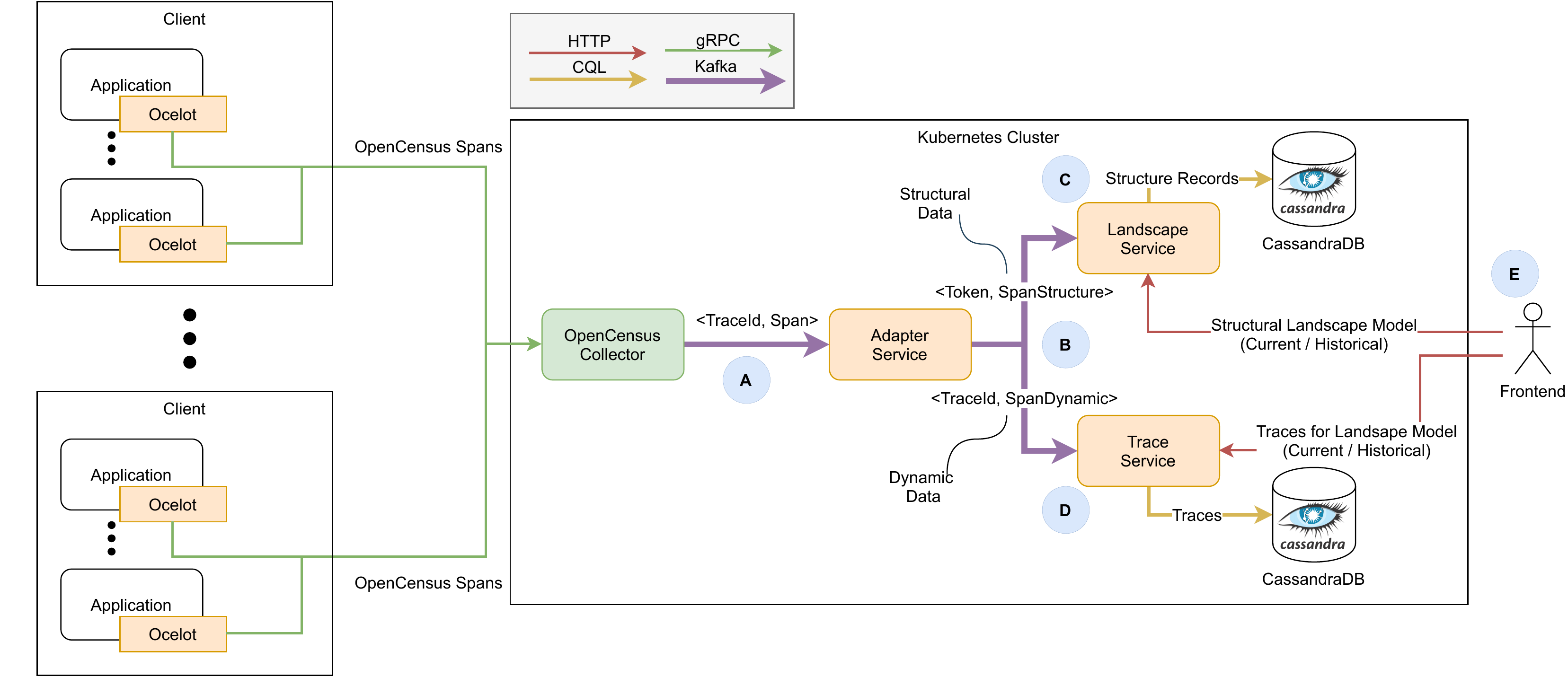}
	\caption{Simplified architectural overview of ExplorViz.}
	\label{fig:explorviz-architecture}
\end{figure*}

The heat map approach by Benomar et al. applies a color to the region below a class based on statically calculated scores for different metrics.
The employed metrics are either concerned with code contribution or the runtime behavior based on execution traces.
For code contributions, extracted and parsed logs from Apache Subversion are used to generate corresponding events.
Events carry information about the changes an author applied to a class, including the importance of the change, its size, and time of change or associated software version.
The changes to files can be filtered within the visualization by their type: all changes, additions, modifications, and removals.
For execution traces, the authors use a custom C-based agent to extract entry and exit events of called operations during the runtime for a given software application.
The gathered data is processed to retrieve call graphs that contain data about the retrieved operation calls, including the order of execution within an execution thread.
The resulting traces are filtered and mapped to the executed use cases of the analyzed program.
Finally, heat maps are calculated whereby the occurrences of a class within a trace correlates with its assigned heat map colors.

Overall, the heat map approach of Benomar et al. shares several similarities with our approach, mostly concerning the chosen visualization. 
However, our visualization differs in the live representation of dynamic data.
In ExplorViz, operation calls between classes are live aggregated and depicted without the need for a heat map, whereas VERSO does only visualize packages and classes.
Another difference lies in the use of live trace analysis.
This enables our heat map to display changes in the runtime behavior of the instrumented program in near realtime. 
Additionally, we also make use of heat maps in other visualization approaches within ExplorViz.
Namely, we include heat map features in our extension for AR and an upcoming integration into our VR extension.
\section{Live Visualization of Heat Maps for Software Cities}\label{sec:main}

\subsection{ExplorViz}\label{sec:main:explorviz}
ExplorViz uses the city metaphor to visualize live runtime behavior based on execution traces.
For that, we continuously monitor executed operations during runtime of a target application with the inspectIT Ocelot Java agent.\footnote{https://inspectit.rocks}
Fig.~\ref{fig:explorviz-architecture} presents a simplified architectural overview of ExplorViz.

\begin{figure*}[!b]
	\centering 
	\subfigure[Heat map indicates high count of created objects for the class \texttt{BaseEntity}.]{\label{fig:explorviz-heat-map}\includegraphics[width=.942\columnwidth]{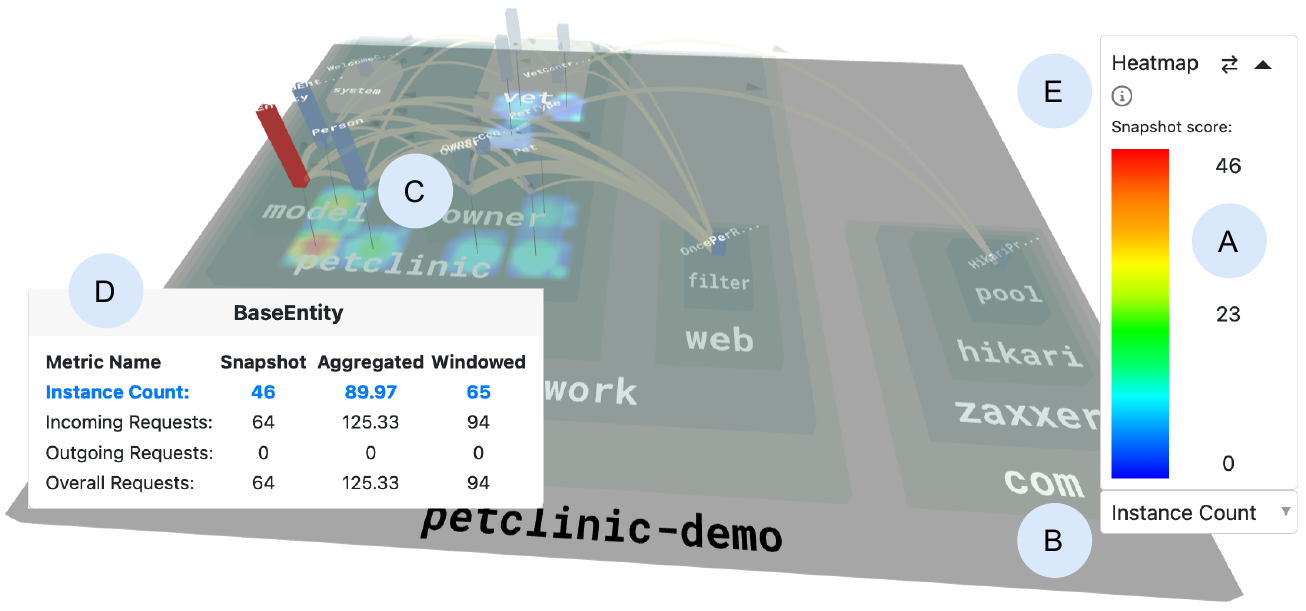}}
	\includegraphics{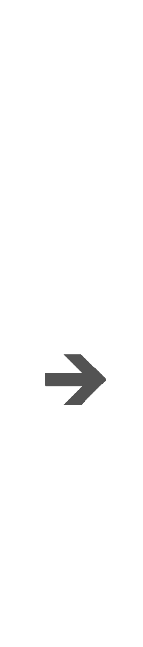}
	\subfigure[Default visualization shows that 20 constructor calls were initiated by the class \texttt{Person}.]{\label{fig:explorviz-popup}\includegraphics[width=.942\columnwidth]{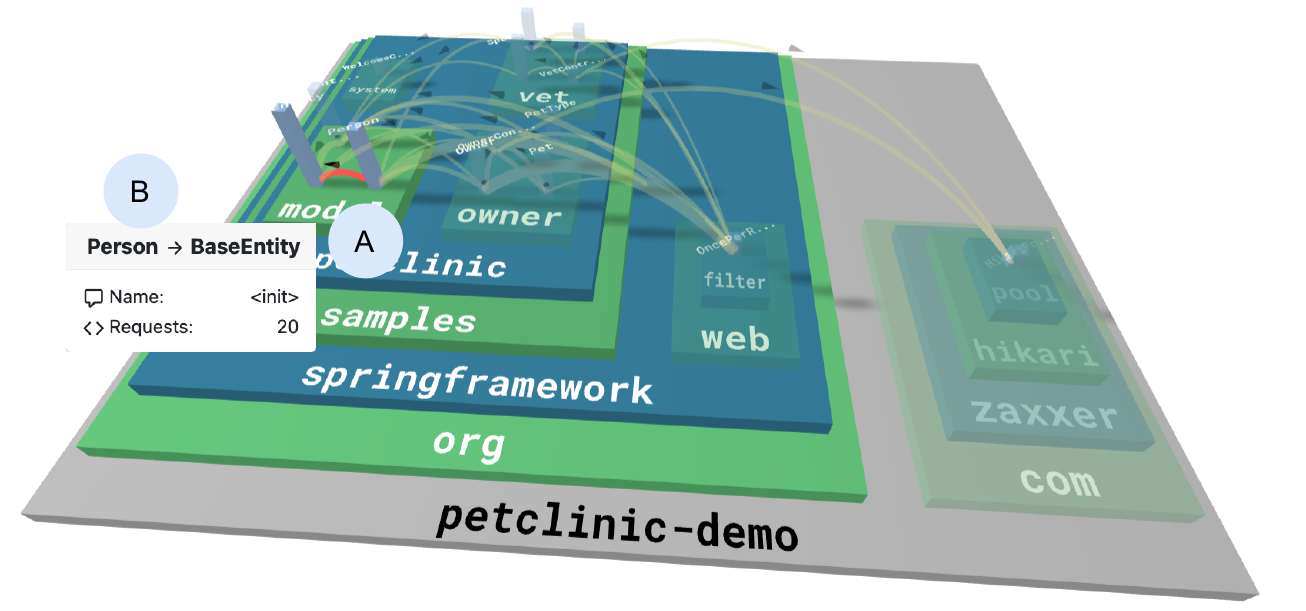}}
	\caption{ExplorViz's visualization with enabled and disabled heat map for the Spring PetClinic.\protect\footnotemark~Popups show additional information.}
	\label{fig:explorviz-heat-map-toggle}
\end{figure*}

Each incoming data set about an executed operation (Fig.~\ref{fig:explorviz-architecture}-A) is separated into structural and dynamic data, and then live processed by ExplorViz's analysis  (Fig.~\ref{fig:explorviz-architecture}-B).
The structural date  (Fig.~\ref{fig:explorviz-architecture}-C) contains the hostname, application name, and fully qualified operation name starting at the top level source code package.
The dynamic date  (Fig.~\ref{fig:explorviz-architecture}-D) contains the timing information of an operation.
This data is used to reconstruct execution traces of the monitored application.
A trace consists of multiple dynamic data entries.
Each instance of a dynamic data entry can be mapped onto a specific instance of a structural data entry.
This partitioning of monitoring data improves the performance.
Structural information is often repeated during execution, since operations are repeatedly called.
Therefore, in high-load situations with multiple concurrent users and monitored applications we can reduce the data volume this way.

ExplorViz's frontend  (Fig.~\ref{fig:explorviz-architecture}-E) updates the depicted software city every few seconds, incorporating all traces and potential new structural information of these past seconds in a single snapshot.
The default value of the update loop is 10 seconds, since in testing we found this value to be suitable.
In our experience, a faster rendering of changing visualization elements, e.g., the height of classes, impairs the user's understanding of the visualization.

\subsection{Heat map visualization}\label{sec:main:heatmap}
ExplorViz's heat map mode is implemented within the frontend component (Fig.~\ref{fig:explorviz-architecture}-E).
Since it is an extension to our existing visualization, the implementation uses the same data that is processed by ExplorViz's backend components.
Therefore, the frontend's update loop also triggers recalculations of software metric scores for the heat map, next to updating the software city.
The calculated scores provide the basis of our heat map visualization.
They, together with the heat map visualization, are novel contributions of this paper.

\newpage

We chose different class-level dynamic coupling metrics~\cite{geetika2014}, since they suit ExplorViz's dynamic analysis of software systems's runtime behavior.
The heat map visualizes a live calculated metric score.
To enhance usability, users can switch the depicted metric at any time in a frontend dialog.
We also tested color weaving~\cite{benomar2013} to render two different metric scores at the same time.
However, this feature was removed since many external factors affected its visibility and usability, e.g., tools that change the color spectrum (night mode) and small displays (for AR).
Our implementation includes four metrics:

\begin{itemize}
	\item The instance count metric summarizes the number of created objects of a given class.
	\item The import coupling metric (IC\_CD)~\cite{arisholm2004} aggregates the total number of operation calls that were initiated by any object of a given class.
	\item The export coupling metric (EC\_CD)~\cite{arisholm2004} aggregates the total number of operation calls that were received by any object of a given class.
	\item The import \& export coupling metric is a combination of both IC\_CD and EC\_CD. 
	It calculates the overall number of operation calls that were sent and received by any object of a given class.
\end{itemize}

The heat map mode can be extended with custom metrics.
For that, developers only need to implement their metric calculation as JavaScript function in a provided web worker.
The function must return a JavaScript object that includes information such as the metric name and the calculated value.
The precise specification for the return value can be seen in the source code.

Fig.~\ref{fig:explorviz-heat-map-toggle} depicts ExplorViz software city visualization with enabled and disabled heat map overlay.
Tiles represent source code packages.
They can be opened or closed, showing or hiding package internals, respectively.
Tall boxes visualize classes, whereas the height indicates the instance count in the current snapshot.
Operation calls between classes are visualized by the orange communication lines.
A (not pictured) trace replayer highlights cohesive communication lines.
As a result, users can examine which operation calls and classes belong to a specific trace.
\footnotetext{https://github.com/spring-projects/spring-petclinic}

\begin{figure*}[b]
	\centering 
	\includegraphics[scale=1.3]{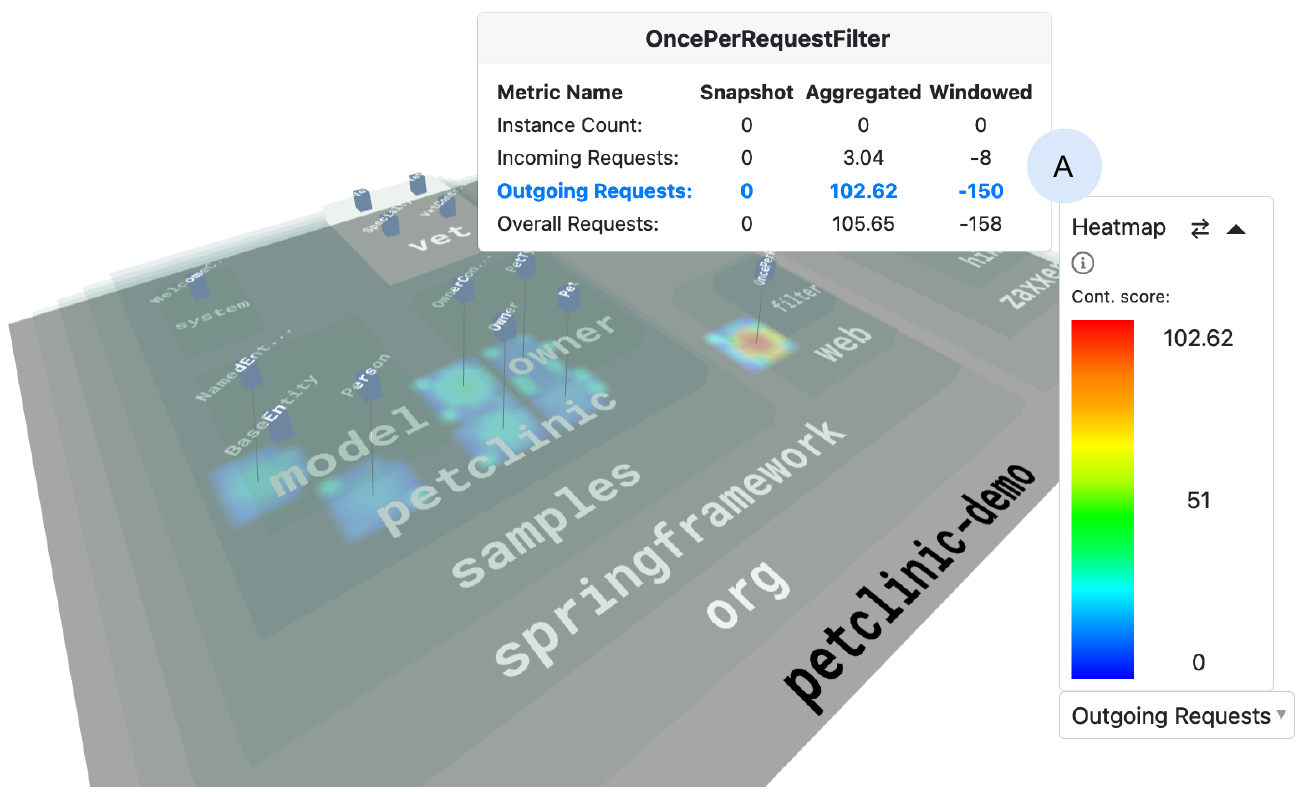}
	\caption{Heat map overlay showing the continuously aggregated metric scores for outgoing communication.}
	\label{fig:explorviz-heat-map-outgoing-communication}
\end{figure*}

Fig.~\ref{fig:explorviz-heat-map} shows a snapshot of ExplorViz visualizing the runtime behavior of the Spring PetClinic.\footnotemark[3]
ExplorViz's heat map is visualized as overlay on top of the application's foundation plate.
A small black line connects visualized classes to their heat map area.
This helps to identify which class belongs to which heat map spot in the three dimensional space. 
We decided against coloring any city metaphor element for the heat map mode, but to use a mostly non-intrusive approach.
This is due to the fact that we think many different colors for the classes might distract users.
However, future research should evaluate how different colorings of city metaphor elements affects the comprehension in comparison to a non-intrusive approach as shown in this paper.

To further enhance visibility, the heat map changes the transparency of the software city.
As a result, it foregrounds the city visualization, but does not neglect its effect.
All software city elements are still visible, while they do not cover the heat map.
The heat map legend (Fig.~\ref{fig:explorviz-heat-map}-A) shows how the color gradient is mapped to the current score of the selected metric.
We followed Röthlisberger's et al.~\cite{Rothlisberger2009} color scheme.
It ranges from blue (cold) to red (hot).
The mapping of metric scores to the color gradient changes every ten seconds due to the frontend's update loop.
Metrics can be switched by using the selection dialog (Fig.~\ref{fig:explorviz-heat-map}-B).

In detail, the heat map in Fig.~\ref{fig:explorviz-heat-map}-C visualizes the number of created objects for all classes of this snapshot during the execution of a use case inside the monitored Spring PetClinic.
Three classes show a high amount of created objects in comparison to the remaining classes.
Since this information is inaccurate, we combined ExplorViz interaction system with the heat map overlay.
As a result, the overall software city visualization provides more information compared to a non-interactive heat map on its own.
\newpage \noindent For example, ExplorViz shows more information for a visualization element when users hover this element with their mouse.
In this case, the resulting popup (Fig.~\ref{fig:explorviz-heat-map}-D) shows that 46 objects were instantiated for the class \texttt{BaseEntity} in this snapshot.

We further investigate which classes instantiated the selected \texttt{BaseEntity} class.
By clicking on this class, we can select this visualization entity (highlighted in red in Fig.~\ref{fig:explorviz-heat-map}) and fade out elements which do not directly communicate with the selected \texttt{BaseEntity} class.
We see that two classes are not faded out, therefore directly communicate with the \texttt{BaseEntity} class.
To achieve a better visibility of communication lines, we disable the heat map overlay. Fig.~\ref{fig:explorviz-popup} also shows an additional feature of ExplorViz's interaction system.
In this case, we selected the communication line between the classes \texttt{BaseEntity} and \texttt{Person} (Fig.~\ref{fig:explorviz-popup}-A).
By hovering the communication line with the mouse cursor, we can again see precise information about this operation call (Fig.~\ref{fig:explorviz-popup}-B).
Here, we see that class \texttt{Person} called the constructor of class \texttt{BaseEntity} 24 times.
This behavior hints at an inheritance relationship or object composition.
A source code lookup confirmed the first assumption.

Fig.~\ref{fig:explorviz-heat-map}-E hints another feature of ExplorViz's heat map overlay.
Depending on the program comprehension task to solve, it might be useful to incorporate previous snapshots, i.e., previous metric scores, into the heat map visualization.
The following section explains how we take time into account.

\subsection{Considering time}\label{sec:main:time}
Next to the metric that is visualized, users can also select if ExplorViz's heat map should incorporate previous metric scores.
To achieve this, we provide three interchangeable heat map modes, which users can select by clicking on the two arrows in the heat map legend (\ref{fig:explorviz-heat-map}-E).

\paragraph{Snapshot mode}%
The heat map only considers the data of the current snapshot, i.e., data that is currently visualized by the software city.
This mode highlights information that can also be found when analyzing the software city, such as the amount of incoming and / or outgoing communication of a class within the snapshot.
We think that this mode has the potential to become a great supplement to the software city visualization.
Users can easier find their desired information based on the metric they choose, e.g., IC\_CD.

\paragraph{Continuously aggregated mode}%
In this mode, each score of the selected metric is continuously aggregated.
Therefore, the heat map incorporates any past runtime information.
If there is no previous metric score, as it is the case with the beginning of the measurement, the value is taken on its own for the score.
Otherwise, 50\% of the previous calculated metric score is added to the newly computed score.
A useful metric to choose in this mode is the number of outgoing requests, i.e., method calls of other objects, for a given class.
This can be seen in Fig.~\ref{fig:explorviz-heat-map-outgoing-communication}.
The \textit{Aggregated} row shows the continuously aggregated metric score for the class \texttt{OncePerRequestFilter}.
\newpage \noindent The heat map shows that this class has the biggest score for the selected metric since the beginning of record.
Other classes have a score around 20 at most or did not communicate at all.
This fact and its name hint that this class processes a high amount or even all requests that the overall application receives.

\paragraph{Windowed mode}%
In contrast to the continuously aggregated mode, the windowed mode compares the latest snapshot with a previous one.
The distance in time of these snapshots is defined by a time window, which gives this strategy the name.
Its default size is 10.
As a result, the heat map visualizes the comparison of metric scores of the latest snapshot and the ones from the snapshot 10 time units ago.
These scores range from negative to positive.
Negative values represent a decline for the selected metric, while positive values represent a growth with respect to the latest snapshot. 
For example, if a class was instantiated 30 times in the snapshot that is selected by the time window and 20 times in the latest snapshot, the instance count metric score would produce a value of -10.
Fig.~\ref{fig:explorviz-heat-map-outgoing-communication}-A shows this decline for another metric, i.e., outgoing requests for a given class.
The value \\ -150 reveals that the number of outgoing requests of the class \texttt{OncePerRequestFilter} decreased, in comparison to the point in time, which was selected by the time window.

\section{Conclusions \& Future Work}\label{sec:conclusions}
In this paper, we presented an approach  inspired by Benomar et al.~\cite{benomar2013} for live visualization of dynamic software cities with heat map overlays.
We extended our web-based live trace visualization tool ExplorViz, such that users can toggle the heat map and switch their visualized metric at any time.
In contrast to preliminary work, our heat map calculations are based on live trace analysis.
This avoids unnecessary semi-automatic or manual processing steps.
Furthermore, we also render operation calls of an application's runtime as communication lines in the software city visualization.
While the heat map indicates for example a high amount of called operations for a given class, the rendered communication lines allow users to investigate the actual runtime behavior.
This extension to the 3D city metaphor seamlessly integrates in our software visualization tool, such that the heat map overlay is also available in our AR environment and soon usable in VR as well.

Currently, the heat map calculations are performed in ExplorViz's frontend, therefore, on the user's device.
For future work, we will examine if and how a backend service might enhance the performance by overtaking the software metrics calculation.
As with our existing backend services, this service will also be developed as Cloud-native application, taking scalability into account to handle different load~\cite{Fittkau2015Cloud}.
Furthermore, this service might also be used to persist metric scores for a long time.

Regarding the heat map visualization, we will investigate new software metrics for runtime information.
In addition, we want to introduce more interaction features.
One example is the highlighting of all classes belonging to a certain color value.
This might be triggered when a user clicks a specific class.
Adding this feature might prevent users from estimating metric scores and allow them to receive precise information.

\bibliographystyle{IEEEtran}

\end{document}